\documentclass[osajnl,twocolumn,showpacs,showkeys,amsmath,amssymb,citeautoscript,10pt,floatfix]{revtex4-1} %% use 11pt for Applied Optics

\usepackage{color,graphicx,bm}
\setcitestyle{super}

\newcommand{\dlangle}{\left\langle\!\left\langle}
\newcommand{\drangle}{\right\rangle\!\right\rangle}

\begin{document}

\title{Optical Pulse Dynamics in Active Metamaterials with Positive and Negative Refractive Index}                      

\author{Alexander O. Korotkevich}\email{alexkor@math.unm.edu}\affiliation{Department of Mathematics and Statistics, MSC01 1115, 1 University of New Mexico, Albuquerque, NM 87131-0001, USA\\ and \\ 
Landau Institute for Theoretical Physics, Kosygin Str. 2, Moscow, 119334, Russia}
\author{Kathryn E. Rasmussen}\affiliation{Mathematics Department, Brevard College, One Brevard College Dr., Brevard NC 28712, USA}
\author{Gregor Kova\v{c}i\v{c}}\author{Victor Roytburd}\affiliation{Mathematical Sciences Department, Rensselaer Polytechnic Institute, 110 8th St., Troy, NY 12180, USA}
\author{Andrei I. Maimistov}\affiliation{Moscow Institute of Physics and Technology, Dolgoprudnyi, Moscow region, 141700, Russia,\\
and\\
National Research Nuclear University MEPHI, Kashirskoe Ave. 31, Moscow, 115409, Russia}
\author{Ildar R. Gabitov}\affiliation{Department of Mathematics, The University of Arizona,
617 N. Santa Rita Ave., P.O. Box 210089,  Tucson, AZ 85721-0089, USA}

\begin{abstract}
We study numerically the propagation of two-color light pulses through a metamaterial doped with active atoms such that the carrier frequencies of the pulses are in resonance with two atomic transitions in the $\Lambda$ configuration and that one color propagates in the regime of positive refraction and the other in the regime of negative refraction.   In such a metamaterial, one resonant color of light propagates with positive and the other with negative group velocity.    We investigate nonlinear interaction of these forward- and backward-propagating waves, and find self-trapped waves, counter-propagating radiation waves, and hot spots of medium excitation.
\end{abstract}


\maketitle

\section{INTRODUCTION}
 
Recent advances in materials technology have led to the emergence of a new branch of science, the physics of metamaterials. These are artificially engineered composites that exhibit properties unattainable in nature and furnish exciting potential new tools for future technologies~\cite{MMCE:MMCE20634}.   
Electrodynamics of metamaterials
has opened avenues for manipulating light in unforeseen ways, as well as presented new paradigms in basic science~\cite{schurig06,cai06,zhel12}.    Metamaterials exhibiting negative
refractive index within certain frequency ranges~\cite{smith00,shelby01,smith04,shalaev05,zhang05,shalaev07}
provide a particularly promising example,   as they
give rise to a number of unusual optical properties impossible to observe in natural materials.    In addition to negative refraction, i.e., reverse Snell's law~\cite{veselago68}, these properties include perfect lensing~\cite{pendry00,podolskiy05}, reverse Doppler shift~\cite{veselago68}, reverse Cherenkov radiation~\cite{veselago68,lu03}, and, in particular, phase and group velocities pointing in the opposite directions~\cite{agranovich04}.  
Negative refractive index was first studied as a purely hypothetical concept  and analyzed using simple mathematical models~\cite
{mandelshtam45,veselago67,veselago68}.  Only relatively recently was this
concept realized experimentally using metamaterials, first for microwaves~\cite{smith00}, and 
subsequently for infrared and visible light~\cite{shalaev05,zhang05}.   

Nonlinearity in the interaction of light with metamaterials can arise both from the host material and/or from the embedded structures~\cite
{zharov03,agranovich04,scalora05,lazarides05,popov06,gabitov06,litchinitser07b,litchinitser07c,maimistov07a}.  Particularly
intriguing situations arise on the boundaries of negative- and
positive-refractive-index regimes in metamaterials, be it in the physical or the frequency domain. The former include, for example, nonlinear
optical vortices~\cite{shadrivov04} and bistability~\cite{litchinitser07b}.
For the latter, let us recall that negative refractive index exists only inside limited frequency ranges, while outside these ranges the refractive index is positive.
Nonlinearity can facilitate energy interchange among different frequency
components of optical pulses traveling through a nonlinear metamaterial. Two possible scenarios can occur: pulses with broad spectra whose wings belong to regions of different refractive-index sign, or
multi-wave interaction in which the individual wave frequencies belong to
such regions~\cite{popov06,agranovich04,gabitov06,gabitov07b}.  One way to achieve this latter scenario is to dope a metamaterial with active atoms, and launch into this metamaterial two-color light pulses that resonate with two atomic transitions from a pair of energetically-lower level to a common upper level, with one transition frequency lying in the regime of positive and the other in the regime of negative refraction.   This transition configuration is known as the $\Lambda$ configuration~\cite{konopnicki81}.  Likewise, in a chiral metamaterial~\cite{Pendry19112004,PhysRevE.69.026602}, the opposite circular polarizations of a monochromatic light pulse that travel in the opposite regimes of the refractive-index sign can be coupled via the interaction with active atoms through a pair of transitions in the $\Lambda$ configuration.

In this paper, we study two-color light pulses propagating through metamaterials containing embedded active atoms with two electron transitions in the $\Lambda$ configuration, corresponding to the two colors,  for which one  transition frequency lies in the positive-refraction and the other in the negative-refraction regime.    This physical setup provides a perfect example of nonlinear interaction between wave pairs propagating within the two refraction regimes in the metamaterial.     Since the group velocities of such wave pairs have the opposite signs, it is not  surprising to find  that resonantly-interacting waves typically travel in the opposite directions.   This unusual interaction gives rise to a number of effects.     
The first effect we find is self-trapping, resulting in a pair of co-propagating solitary-wave-type pulses similar to the self-trapped pulses in metamaterials exhibiting second-harmonic generation~\cite{Maimistov07,MaimistovGabitovOS08}.  The second effect is the counter-propagation of the linear waves shed by the incident pulse, traveling at the two respective carrier frequencies.    The third effect is the stopping of a pulse as it switches energy from one color to the other,  accompanied by a hot spot of medium excitation left in the metamaterial at the location of the color switching and direction reversal.

Several known mechanisms currently exist  that produce a negative refractive index, such as chirality~\cite{Pendry19112004,PhysRevE.69.026602}, hyperbolic materials~\cite{Podolskiy05b,Naik05062012}, and metamaterials based on the magnetic response utilizing plasmonic resonance~\cite{smith00,shalaev05,zhang05}.    All these approaches yield the same ultimate effect, and so, with no loss of generality, we here focus on the last one.   
We begin with Maxwell's equations for the electric and magnetic fields, coupled to material equations describing the magnetic and dielectric response of the nanostructures and the additional response of the active atoms in the $\Lambda$ configuration.
We restrict our study to coherent phenomena, i.e., the case in which pulses are very short compared to all the relaxation time scales.
 Under the assumption that light pulses are short, but still much longer than the individual wavelengths of the light they contain, we have derived a set of Maxwell-Bloch equations that describe the electric-field envelopes and the medium polarizations and level-occupation densities.   While we only consider idealized, lossless propagation, our equations can be readily adapted to include terms describing losses and pumping.   This is important because metamaterials are quite lossy, especially in the negative-refraction frequency range, and doping them with active atoms has in fact been used to successfully  reduce these losses~\cite{xiao10}.

The remainder of the paper is organized as follows.    In Sec.~\ref{sec:mainmod} we present our modeling assumptions and state the basic physical model that we adopt for our investigation.   In addition, we determine the fundamental temporal and spatial scales in this model.   In Sec.~\ref{sec:mbes}, we sketch a derivation from this model of the Maxwell-Bloch equations for the envelopes of the electric-field components and material variables, which we use in our numerical simulations.     In Sec.~\ref{sec:setup}, we discuss the physical setup underlying our numerical simulations.   In particular, we discuss the choice of the parameters, as well as the initial and boundary conditions.   In Secs.~\ref{sec:coprop} and \ref{sec:hot}, we discuss the phenomena we have uncovered in our model: co-propagating pairs of self-trapped pulses, counter-propagation of radiation components, and hot-spot formation in the medium.   Finally, in Sec.~\ref{sec:conc}, we present our conclusions.

\section{MODEL\label{sec:model}}

To set the stage for our numerical investigation, in this section, we briefly outline a derivation of the envelope equations that describe the propagation of optical pulses in a $\Lambda$-configuration metamaterial with embedded 
nano-inclusions.  Ultimately, we will assume our model to describe a metamaterial fabricated in such a way that one of the two allowed atomic-transition frequencies takes place in a regime of negative refractive index  and the other transition 
frequency in a  regime of positive refractive index.  Our goal is to investigate the interaction among waves propagating 
in the two regimes, and the resulting dynamical phenomena emerging from this interaction.

\subsection{Fundamental Equations and Metamaterial  Model \label{sec:mainmod}}

Electromagnetic field propagation through a dielectric (meta)material in the absence of  macroscopic currents and free charges is described by 
Maxwell's Equations
\begin{subequations}\label{maxwell}
\begin{equation} \label{bud} \nabla \times \mathbf{E} = -\frac{1}{c}\frac{\partial \mathbf{B}}{\partial t} ,  
\qquad \nabla \times \mathbf{H} = 
 \frac{1}{c}\frac{\partial \mathbf{D}}
{\partial t} , \end{equation}
\begin{equation} \label{great} \nabla \cdot \mathbf{D} = 0,
\qquad \nabla \cdot \mathbf{B} = 0,  \end{equation} 
\end{subequations}
where $\mathbf{E}$ represents the electric field,  $\mathbf{B}$ the magnetic induction, $\mathbf{H}$ the magnetic field, and
$\mathbf{D}$ the electric displacement field. 
The constant $c$ is the speed of light in vacuum.  
 The interaction between the electromagnetic field  and the (meta)material doped with active atoms is expressed through the constitutive relations
\begin{subequations}
\begin{equation} \label{sean} 
\mathbf{B}  = \mathbf{H}+4\pi\mathbf{M}, \qquad
\mathbf{D}  = \mathbf{E}+4\pi\mathbf{P}+4\pi\mathbf{Q},
\end{equation}
for the magnetization  $\mathbf{M}$ and polarization $\mathbf{P}$ due to the nanoinclusions and the polarization  $\mathbf{Q}$ due to the active atoms.

The dynamics of the magnetization $\mathbf{M}$ and polarization $\mathbf{P}$
are modeled by the oscillator equations \cite{allen87,ziolkowski,gabkenmaim10}  
\begin{equation} \label{lynn} \frac{\partial^2 \mathbf{P}}{\partial t^2}+\omega_{P}^2 \mathbf{P}= \frac{\omega_{P}^2 \gamma}{4\pi} \mathbf{E},
\end{equation}
\begin{equation} \label{kathy} \frac{\partial^2 \mathbf{M}}{\partial t^2}+\omega_{M}^2 \mathbf{M}= -\frac{\beta}{4\pi} \frac{\partial^2 
\mathbf{H}}{\partial t^2}.  
\end{equation}
Here, $\omega_P$ is the plasmonic oscillation frequency and $\omega_M$ is the magnetic 
resonance frequency associated with the inclusions, and $\gamma$ and $\beta$ are form factors specific for the given inclusion type and material properties of the surrounding dielectric in the metamaterial, which can be evaluated by careful homogenization~\cite{panina02,smith06}, or else obtained from the measured plasmonic and magnetic resonance
curves~\cite{klar06}.  
Equation (\ref{lynn}) is the classical Lorentz model \cite{allen87} and (\ref{kathy}) is its analog in a magnetic material~\cite{Pendry99,PhysRevE.65.036622,gabitov06}.

The polarization contribution $\mathbf{ Q}$ due to the active atoms is modeled as 
\begin{equation} \label{jon} \mathbf{Q}=\mathcal{N}  \sum_{\stackrel{l,j=1}{l \ne j}}^3  \dlangle a_l^{*}a_j \drangle \cdot \mathbf{r}_{lj}  , 
\quad  \mathbf{r}_{lj}=\mathbf{r}_{jl}^*, \quad \mathbf{r}_{ll}=0, \end{equation}
where $ a_l $, $l=1, 2, 3$, are the quantum amplitudes of the three active
atomic levels in each individual atom, $\mathcal{N}$ is the density of the active atoms in the material, $\mathbf{r}_{lj}$ 
are the electric dipole moments of the transitions between the $l$-th and $j$-th state, and $\dlangle \cdot\drangle$ denotes averaging over all the atoms in an infinitesimally small volume at the location $z$.  For every atom, the amplitudes $a_m$ satisfy the equation
\begin{equation} \label{mike} i \hbar \frac{\partial}{\partial t} a_m = \hbar \omega_m a_m - \displaystyle\sum_{n=1}^3 \mathbf{E} 
\cdot \mathbf{r}_{mn}a_n , \end{equation}
\end{subequations}
for $m=1, 2, 3$, 
which can be derived from Schroedinger's equation \cite{cohen}.  
Here $\hbar$ is Planck's constant and $\hbar\omega_m$ are the energies of the three working levels in the active atoms embedded in the metamaterial, with $\omega_m$ being the corresponding quantum frequencies.  In what is to follow, we will assume that they are narrowly distributed around three fixed values.   

For active atoms that have a pair of transitions in the $\Lambda$ configuration, we  assume that the frequency $\omega_1$ corresponds to the excited energy level, and $\omega_2$ and $\omega_3$ to the two lower energy levels.   In this configuration, the transition between the two lower levels is forbidden, and so the corresponding  
electric dipole moments vanish,
\begin{equation}\label{excellent}
\mathbf{r}_{23}=\mathbf{r}_{32}=0.
\end{equation}

\subsection{Maxwell-Bloch Envelope Equations\label{sec:mbes}}

We assume wave propagation in the medium to be one-dimensional, varying in the $z$-direction only, so that the fields $\mathbf{E}$, $\mathbf{B}$, $ \mathbf{P}$, $ \mathbf{M}$, $ 
\mathbf{D}$,  and $\mathbf{H}$ all lie in the $xy$-plane.  
Equations (\ref{great}) are then satisfied automatically.  
We  non-dimensionalize equations (\ref{maxwell}), 
(\ref{lynn}), (\ref{kathy}), (\ref{jon}), and (\ref{mike}) using a number of  fundamental scales in the system.  We let $\omega_0=\max_{j\neq k} |\omega_j-\omega_k|$ denote the typical size of the quantum frequencies $\omega_n$ in Eq.~(\ref{mike}), which is inversely proportional to the wavelengths of the light absorbed and emitted in the corresponding atomic transitions.  We let $d= \max{ |\mathbf{r}_{mn}|}$ denote the typical electric dipole strength of the electron transitions. 
We define the \emph{cooperative frequency}
\begin{equation}
\omega_c=\sqrt{\frac{2\pi \mathcal{N} d^2 \omega_0}{\hbar}},
\end{equation}
which is the typical response scale of the active atoms in the medium, while its reciprocal is proportional to the temporal width of the light pulses involved in the  interaction with the metamaterial.    
Using the scales $\omega_0$, $\omega_c$, and $d$, we rescale the spatial and temporal variables as 
${\omega_c z}/{c}\to z$, $\omega_c t\to t$,
the frequencies as $\omega/\omega_0\to\omega$, 
$\Omega_P={\omega_P}/{\omega_0}$, $\Omega_M= {\omega_M}/{\omega_0}$, $\Omega_m = {\omega_m}/{\omega_0}$, for $m=1,2,3$,
the electromagnetic-field and material variables as
$ {d  \mathbf{E} }/{\hbar\omega_c}\to\mathbf{E}$, 
${d  \mathbf{H} }/{\hbar\omega_c} \to\mathbf{H}$, 
$ {4\pi d  \mathbf{P} }/{\hbar\omega_c} \to\mathbf{P}$, 
$ {4\pi d  \mathbf{M} }/{\hbar\omega_c} \to\mathbf{M}$, and the electric dipole moments as $ \bm{\alpha}_{lj}= \mathbf{r}_{lj}/d$.

We restrict our study to the case of the slowly-varying envelope approximation when light pulses include many electric-field oscillations, i.e., $\omega_c\ll \omega_0$. 
We  thus assume the electric field to be a sum of slowly-varying modulated plane waves 
\begin{equation}  \label{dodo}
\mathbf{E} (z,t) = \mathbf{E}_+(z, t )e^{i\theta_+} + \mathbf{E}_-( z,  t )e^{i\theta_-} + \mbox{c.c.},
\end{equation}
where c.c. stands for  complex conjugate, $\omega_\pm$ are the carrier frequencies of the electric fields, and the fast phases are described by  the expressions
\begin{equation}\label{phases}
\theta_{\pm}=\frac{\omega_0}{\omega_c}\left(k_{\pm}z-\omega_{\pm}t\right) .
 \end{equation}
We assume the  
fields $\mathbf{H}$, $\mathbf{P}$, and $\mathbf{M}$ to behave in a like 
manner.    
Here, the dimensionless  wavenumbers and frequencies satisfy the usual dispersion relation
\begin{subequations}
\begin{equation} \label{will}
k^2(\omega) =\omega^2\mu(\omega)\epsilon(\omega), 
\end{equation}
where $\epsilon(\omega)$ and $\mu(\omega)$ are the dielectric permittivity and magnetic permeability, respectively, given by the expressions
\begin{equation}\label{alaska}
\epsilon(\omega) = 1 + \frac{\gamma \Omega_P^2}{\Omega_P^2 - \omega^2} , \qquad
\mu(\omega) = 1 + \frac{\beta \omega^2}{\Omega_M^2 - \omega^2} ,
\end{equation} 
\end{subequations}
which follow from the oscillator equations (\ref{lynn}) describing the medium polarization  $\mathbf{P}$ and magnetization $\mathbf{M}$.

For the polarization $\mathbf{Q}$ due to the active atoms, defined in Eq.~(\ref{jon}),   we assume the form
\begin{align}
\mathbf{Q} =& - i \int_{-\infty}^\infty \left[ \rho_+ (z,t,\nu) \bm{\alpha}_+^* e^{i\theta_+}  \right.\nonumber \\ & \qquad\qquad \left. + \rho_- (z,t,\nu) \bm{\alpha}_-^*e^{i\theta_-}  -  \mbox{c.c.}\right]g(\nu) \,d\nu,
\end{align}
where $\bm{\alpha}_+ = \bm{\alpha}_{12}$ and $\bm{\alpha}_- = \bm{\alpha}_{13}$, and the components  $\rho_\pm (z,t,\nu)$ are detuned from the exact resonance with the carrier frequencies $\omega_\pm$ of the electric-field components $\mathbf{E}_\pm(z,t)$ by the amount $(\omega_c/\omega_0)\nu$.   In other words, we assume that the respective atomic transition frequencies $\Omega_1-\Omega_2$ and $\Omega_1-\Omega_3$ are related to $\omega_+$ and $\omega_-$ by the relations $\omega_+ = \Omega_1 - \Omega_2+(\omega_c/\omega_0) \nu$ and $\omega_- = \Omega_1 - \Omega_3+(\omega_c/\omega_0) \nu$.    The detuning $\nu$ arises mainly due to the motion of the active atoms, which gives rise to a Doppler shift in the transition frequencies; $\nu$ is therefore distributed according to a function $g(\nu)$ with $\int_{-\infty}^\infty g(\nu)\,d\nu=1$, which describes the shape of the spectral line~\cite{allen87}.   In assuming that the actual detuning is of size $(\omega_c/\omega_0)\nu$, we assume that this line is very narrow as compared to the typical size of the variations of the 
dielectric permittivity $\epsilon(\omega)$, magnetic permeability $\mu(\omega)$, and the wave vector $k(\omega)$ with the frequency $\omega$.   

The polarization components $\rho_\pm$, corresponding to the two atomic transitions in the $\Lambda$-configuration medium,  are expressed in terms of the atomic level amplitudes $a_k$ as 
%\begin{subequations}
%\begin{align}
$\rho_+ (z,t,\nu) = i\left\langle a_1(z,t) a_2^*(z,t)\right\rangle e^{-i\theta_+}$ and %,\\
$\rho_- (z,t,\nu) = i\left\langle a_1(z,t) a_3^*(z,t)\right\rangle e^{- i\theta_-}$,
%\end{align}
 where $\langle \cdot \rangle$ denotes ensemble averaging over all the atoms in the sample in an infinitesimal volume at the location $z$ and time $t$, whose transition frequencies are detuned by $(\omega_c/\omega_0)\nu$ from the resonance with the frequencies of the respective electric field components. 
Likewise, we express the polarization component due to the forbidden transition between the two lower levels as 
$%\begin{equation}
\rho(z,t,\nu) = \left\langle a_2(z,t) a_3^*(z,t) \right\rangle e^{i (\theta_+-\theta_-)}$,
%\end{equation}
and the level-occupation numbers as 
%\begin{equation} 
$N = \left\langle |a_1|^2  \right\rangle$, %\quad 
$n_+ = \left\langle  |a_2|^2 \right\rangle$, %\quad 
and $n_- = \left\langle  |a_3|^2 \right\rangle$.
%\end{equation}\label{material}
%\end{subequations}

For the electric-field components $E_+ = 2 \mathbf{E}_+ \cdot \bm{\alpha}_+$ and $E_- = 2\mathbf{E}_- \cdot \bm{\alpha}_-$, and the macroscopic polarization and occupation variables, $\rho_\pm$, $\rho$, $N$, and $n_\pm$, % in Eqs.~(\ref{material}),
we derive the set of  Maxwell-Bloch equations
\begin{subequations} \label{skippy} \begin{equation} \label{wood}
\frac{\partial E_\pm}{\partial z} + \frac{1}{v_\pm} \frac{\partial E_\pm}{\partial t} = 
2\omega_\pm \sqrt{\frac{\mu_\pm}{\epsilon_\pm}}\left|\bm{\alpha}_\pm \right|^2  \int_{-\infty}^\infty \rho_\pm (\nu)  g(\nu) \,d\nu ,
\end{equation} 
\begin{equation} \label{rhoplus}
\frac{\partial \rho_+}{\partial t}  = i\nu \rho_+ + \frac{1}{2} [ E_+(N-n_+)-E_-\rho^* ] ,
\end{equation} \begin{equation}\label{rhominus}
\frac{\partial \rho_-}{\partial t}  =  i\nu \rho_- + \frac{1}{2} [ E_-(N -n_-)-E_+ \rho ] ,
\end{equation}
 \begin{equation}\label{rhoeq}
\frac{\partial \rho}{\partial t}  = \frac{1}{2} [ E_+^* \rho_- + E_- \rho_+^* ] . 
 \end{equation}
  \begin{equation}\label{neq}
\frac{\partial N}{\partial t}  =  -\frac{1}{2} [ E_+ \rho_+^* + E_+^* \rho_+ +E_-\rho_-^* +E_-^* \rho_- ] ,
\end{equation} \begin{equation}\label{npmeq}
\frac{\partial n_\pm}{\partial t}  = \frac{1}{2} [ E_\pm \rho_\pm^* + E_\pm^* \rho_\pm ] ,
\end{equation} 
\end{subequations}
where $\mu_\pm= \mu(\omega_\pm)$ and $\epsilon_\pm=\epsilon(\omega_\pm)$, and  the group velocities $v_\pm=v (\omega_\pm)$ are computed from the dispersion relation (\ref{will}) via the formula
$
v(\omega) = 1/k'(\omega)
$, 
with the prime denoting the derivative with respect to the frequency $\omega$, which leads to the expression
\begin{equation} \begin{aligned}
\frac{1}{v_\pm} =&  \frac{\omega_\pm}{2} \sqrt{\left( 1+ \frac{\gamma \Omega_P^2}{\Omega_P^2-\omega_\pm^2} \right) 
\left( 1+ \frac{\beta \omega^2}{\Omega_M^2 - \omega_\pm^2} \right) }  \\ 
& \times \left[ \frac{2}{\omega_\pm} + \frac{\gamma \Omega_P^2}{\left( \Omega_P^2 (1+\gamma ) - \omega_\pm^2 \right) \left( \Omega_P^2 - \omega_\pm^2 \right)} \right. \\  & \left. \quad + 
\frac{\beta\Omega_M^2}{\left(\Omega_M^2 + \omega_\pm^2 (\beta - 1) \right) \left( \Omega_M^2 - \omega_\pm^2 \right) } \right].
\end{aligned}\end{equation}
We will investigate the solutions of equations (\ref{skippy}) numerically in the remainder of this paper.

\section{RESULTS \label{sec:results}}

Our goal in the remainder of this paper is to investigate a doubly-resonant interaction between light pulses and a metamaterial doped with active atoms in the $\Lambda$ configuration, such that one of the two corresponding atomic transitions takes place in the positive and the other in the negative refractive-index regime.  Using numerical simulations, we 
uncover several phenomena that arise as consequences of this interaction.

\subsection{Physical Setup\label{sec:setup}}
\begin{figure}[htp]
\begin{center}
\includegraphics[width=2.25in]{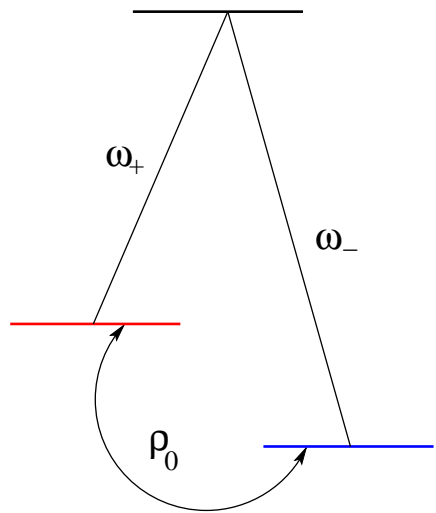}
\end{center}
\caption{$\Lambda$-configuration system with transition frequencies 
$\omega_+$ and $\omega_-$.    The transition with the frequency $\omega_+$ corresponds to the response 
of the metamaterial in a regime with a positive index of refraction, and the transition with the frequency $\omega_-$ corresponds to the response of the metamaterial in a regime with a negative index of 
refraction.   The medium is initially prepared so that a nonzero component of the polarization, $\rho_0$, corresponding to the two-photon transition between the two lower levels, is present.}
\label{lambda2}
\end{figure}

\begin{figure}[htp]
\centering
\includegraphics[width=2.5in]{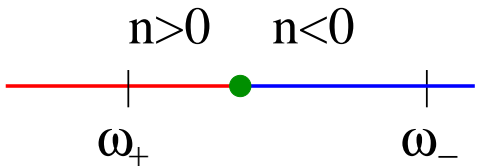}
\caption{Due to the resonant nature of the negative refractive index phenomenon, the sign of the refractive index 
depends on the 
frequency. The intermediate (forbidden) range of frequencies, in which $\epsilon$ and $\mu$ have (real parts with) opposite signs and there is 
no wave propagation, is depicted by the green dot. }
\label{lambda1}
\end{figure}

We simulate light pulses with carrier frequencies which are centered around the 
values  $\omega_+$ and $\omega_-$, as depicted in Fig. \ref{lambda2}, around which the transition 
frequencies of the atoms in the $\Lambda $ configuration embedded in the resonant metamaterial, $\Omega_1-\Omega_2$ and $\Omega_1-\Omega_3$, are also narrowly centered.        In a typical metamaterial, light interacts resonantly with metallic nanoinclusions embedded in it, and due to the 
resonant nature of the negative refraction, the value of the refractive index is negative, $n<0$, only within a limited interval of frequencies 
of $\omega$.  We will study the case when $\omega_+$ is in the regime  of positive refraction, $n>0$, and $\omega_-$ is in the regime of negative  
refraction, $n<0$, see Figure \ref{lambda1}, so that the group velocities of the two pulse components point in opposite directions.  
To best focus on the opposite directionality of the two electric-field-component group velocities $v_\pm$, we consider an idealized version of
equations (\ref{skippy}) in which we set all constants to 1, except $v_\pm$, which we set to $v_\pm=\pm 1$.   This choice may also correspond exactly to the case of a chiral metamaterial, in which the two $\Lambda$-configuration transitions in the embedded active atoms correspond to the same frequency but different polarizations of light~\cite{Pendry19112004,PhysRevE.69.026602}.   In addition, we consider limit of a narrow spectral line, when $g(\nu) =\delta(\nu)$, where $\delta(\cdot)$ is the Dirac delta function.

The physical setup that we consider in our simulations is as follows: Initially, the system is in the ground state, $N(z,0,\nu)=0$, which is 
characterized by a relative distribution of the ground-level populations $n_+(z,0,\nu)$ and $n_-(z,0,\nu)$; the medium polarization due to the one-photon transitions is absent, $\rho_+(z,0,\nu)=\rho_-(z,0,\nu)=0$; the ground states are coupled 
through the initial polarization $\rho(z,0,\nu)=\rho_0(z,\nu)$.  There is no light in the medium initially, $E_+(z,0)=E_-(z,0)=0$, and a right-propagating incident pulse, $E_+(0,t)=E_{+0}(t)$,  is injected at the left edge of the metamaterial sample.  We set $E_-(l,t)=0$, where $l$ is the $z$-coordinate of the right edge of the sample, as we inject no left-propagating electric-field component into the metamaterial.  In the simulations to follow, we take for the boundary condition $E_{+0}(t)$ the Gaussian function given by
\begin{equation}\label{gaussian}
E_{+0}(t)=Ae^{-\left( t -t_d \right)^2 / 2\sigma^2},
\end{equation}
where $A$ is the amplitude, $t_d$ is the delay time, and $\sigma$ the width of the injected pulse.  Specifically, for the latter two parameters, we fix the dimensionless quantities  $t_d=3$ and $\sigma=1$ throughout our investigation.   In addition, we fix the length of the sample to be $l=40$. 

In the majority of previous studies, the initial polarization component $\rho_0(z,\nu)$ was assumed to vanish, which corresponds to the absence of
coherence between the two lower-energy states in the $\Lambda$-configuration medium.     A $\Lambda$-configuration medium with nonzero $\rho_0$ can be prepared by irradiating it with a two-color electromagnetic field in such a way that the transition between the two lower levels is in a two-photon resonance.   This can be achieved if the difference between the frequencies of the two colors of the light used in preparing the medium equals $\omega_+-\omega_-$.   Since a dipole transition between these two lower levels is forbidden, the induced coherence between them, reflected in the polarization component $\rho_0$, will persist much longer than other polarization components.   The nonzero component $\rho_0$ provides an immediate, nonlinear coupling  between the two electric field components $E_{\pm}$, regardless of their size.

\begin{figure} 
\centering
\includegraphics[width=3.3in]{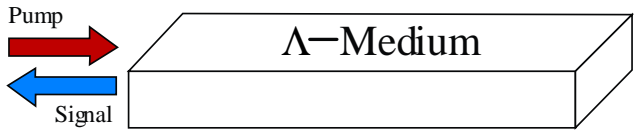}
\caption{A schematic of the physical setup investigated numerically: in this case only the red light, the pump is sent into the 
$\Lambda$-configuration metamaterial.  This red light is partly converted into blue light, which travels 
in the opposite direction.}
\label{setup1}
\end{figure}

We are primarily interested in the physical setup in which a right-traveling electric field component is injected into a metamaterial 
containing active atoms in the $\Lambda$ configuration, 
and a left-traveling electric field component is generated via color switching.  We are also interested in exploring the precise details of how
 this second color of light propagates in the opposite direction of the original input light.  A schematic of this setup 
situation is displayed in Figure ~\ref{setup1}.  Here, the red light,  labeled pump, is injected into the metamaterial and propagates to the right.  The goal is for the 
 red light to be converted into blue light, labeled signal, which would propagate in the negative 
refractive index regime and so to the left, and eventually exit the medium at the 
same end at which the red light entered it.  

We now describe the phenomena we have observed in our numerical simulations, which arise as a consequence of the color conversion discussed above.   

\begin{figure}[htbp]
\includegraphics[width=3.3in]{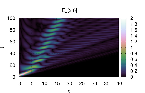}
\includegraphics[width=3.3in]{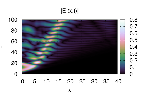}
\caption{\label{rnn.3.8.2} Self-trapped, co-propagating pulse pair. Non-vanishing initial conditions: $\rho_0 = 0.3$, $n_+ =0.8$, $n_- =0.2$.   Amplitude of the injected  Gaussian pulse $E_{+0}(t)$:  $A=2$.}
\end{figure}

\begin{figure*}[t]
\centerline{
\includegraphics[width=2.25in]{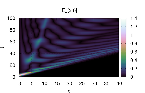}
\includegraphics[width=2.25in]{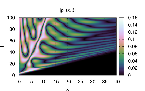}
\includegraphics[width=2.25in]{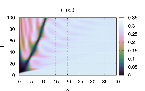}
}
\centerline{
\includegraphics[width=2.25in]{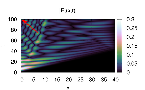}
\includegraphics[width=2.25in]{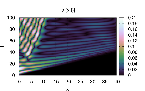}
\includegraphics[width=2.25in]{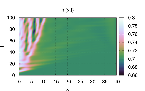}}
\caption{\label{EpAmp1.4} Self-trapped, co-propagating pulse pair and counter-propagating radiation. Non-vanishing initial conditions: $\rho_0 = 0.2$, $n_+ =0.3$, $n_- =0.7$.   Amplitude of the injected  Gaussian pulse $E_{+0}(t)$:  $A=1.4$.}
\end{figure*}

\subsection{Self-Trapping and Counter-Propagation \label{sec:coprop}}

 The results of the simulations we display in Figs.~\ref{rnn.3.8.2} and \ref{EpAmp1.4} show a self-trapped, co-propagating pair of nonlinear pulses with oscillating amplitudes, which forms in both electric-field components, $E_+$ and $E_-$, from the incident pulse with the nonzero component  $E_{+0}(t)$ in Eq.~(\ref{gaussian}).   These components travel simultaneously in the same direction and with the same velocity despite the fact that the linear group velocities $v_\pm$ corresponding to their respective refraction regimes point in the opposite directions.    

For the simulation depicted  in Fig.~\ref{rnn.3.8.2} , the amplitudes $E_+$ and $E_-$ of the self-trapped pulse pair contain virtually all the energy of the incident pulse.  These amplitudes are almost equal, and the resulting trapped pair of pulses appears to propagate at a constant velocity, and considerably more slowly than the group velocity $v_+$ of the positive refraction regime.   The  two co-propagating pulses maintain rather complicated, in-phase, oscillatory shapes during their propagation, and shed a small amount of radiation.     

In Fig.~\ref{EpAmp1.4}, we display a different setup in which the amplitude of the incident pulse $E_{+0}(t)$ and the initial polarization $\rho_0$ are smaller, while the initial population difference has the opposite sign from those used to generate Fig.~\ref{rnn.3.8.2}.
In this case, the emerging  co-propagating
pulse-pair not only propagates more slowly than the linear group velocity $v_+$ of the positive refraction regime, but also 
appears to decelerate during propagation.   Note that the amplitude of the pulse-component $E_-$ shown in in Fig.~\ref{EpAmp1.4}, propagating in the negative refraction regime,  is considerably smaller than that of the pulse-component $E_+$, propagating in the positive refraction regime.    

The spatio-temporal distribution of the electric-field component $E_+$ in Fig.~\ref{EpAmp1.4} shows a forerunner wave, propagating with the group velocity $v_+$  positive refraction regime and with an oscillatory distribution of its tails, followed by the self-trapped pulse, and the waves in the region behind this pulse.   The forerunner wave creates an oscillatory continuous-radiation field in the component $E_-$,  with a much smaller amplitude than the component $E_+$, which both propagate in the region between the forerunner wave and the self-trapped pulse.   Another source of continuous radiation is the self-trapped pulse itself, which largely sheds radiation in the region behind it.    Note that these two regions of continuous radiation exhibit very distinct features:  One can easily notice the radiation in $E_-$ emerge from the peaks in the oscillations of the self-trapped pulse and propagate backward in the region behind this pulse, as indicated by the arrow in Fig.~\ref{EpAmp1.4}.   Note that this radiation still travels more slowly than the group velocity $v_-$  in the negative refraction regime.

From the middle two panels in Fig.~\ref{EpAmp1.4}, one can clearly see the excitation of the medium polarizations $\rho_+$ and $\rho_-$, induced by the co-propagating pulse pair.   A closer look shows that the electric-field and polarization oscillations along this pulse have the opposite phase, indicating that an oscillatory exchange of energy between the electric field and the medium takes place as the pulse propagates.    Note the sharp dip of the polarization component $\rho_+$ along the pulse trajectory, which is where the amplitude of this component changes sign.
In addition, a depletion of the electron population density $n_+$ appears along the pulse trajectory, accompanied  by a corresponding increase in the population density $n_-$.

\subsection{Hot Spot\label{sec:hot}}

\begin{figure*}[htbp]
\centerline{
\includegraphics[width=2.25in]{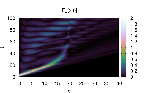}
\includegraphics[width=2.25in]{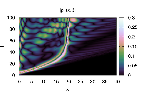}
\includegraphics[width=2.25in]{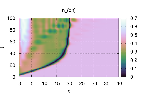}
}
\centerline{
\includegraphics[width=2.25in]{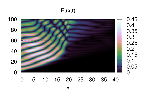}
\includegraphics[width=2.25in]{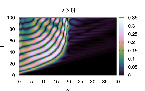}
\includegraphics[width=2.25in]{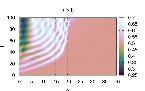}}

\caption{\label{rnn.2.5.5} Hot spot. Non-vanishing initial conditions: $\rho_0 = 0.2$, $n_+ =n_- =0.5$.   Amplitude of the injected  Gaussian pulse $E_{+0}(t)$:  $A=2$.}
\end{figure*}

The incident pulse with the nonzero component $E_+$, displayed in Fig.~\ref{rnn.2.5.5} propagating in the positive-refraction regime,  slows down to a halt and decays in time until it is virtually extinguished.   Part of this pulse switches into the component $E_-$, which contains a great deal of radiation propagating backward.  Note that the amplitude of the incident pulse $E_{+0}(t)$ is the same as in Fig.~\ref{rnn.3.8.2},  the initial polarization $\rho_0$ is the same as in Fig.~\ref{EpAmp1.4}, and the initial population densities are equal.

In addition, we see that at the location at which the entire incident pulse has disintegrated and switched into the component $E_-$, a strong excitation in the medium polarization $\rho_+$ and depletion of the occupation density $n_+$ emerges.   A considerable amount of energy appears to be deposited in the medium at this location.   This energy  forms a stationary hot spot of medium excitation, which also does not appear to noticeably decay in time.   The polarization $\rho_-$ and occupation density $n_-$ appear to also reflect the underlying backward propagating continuous radiation in the $E_-$  component of the electric field.

\section{CONCLUSIONS\label{sec:conc}} 

In this paper we have investigated the nonlinear coupling and consequent flow of energy between light pulses propagating in the positive refractive index regime and light pulses 
propagating in the negative refractive index regime (or vice versa)
  within a resonant metamaterial doped with active atoms in the $\Lambda$ configuration.  
  
The phenomena we have observed in our numerical simulations include the creation of a self-trapped, co-propagating, nonlinear pulse pair that propagates with neither of the group velocities.  In particular, this pulse behaves in an oscillatory fashion and 
propagates in the same direction but with a velocity slower than 
the original incident pulse.  The radiation this pair sheds typically propagates in the directions indicated by the linear group velocities of its respective color.

We have also observed a  pulse with the color corresponding to a positive-index regime to almost completely switch into the negative index regime where radiation is created, which travels backward.  
Simultaneously, a hot spot of permanent medium excitation is formed when the pulse in the positive index regime slows and eventually stops.  From the location of this hot spot, the 
electric field in the negative index regime sheds backward propagating radiation.   The energy deposited in the hot spot appears to remain stationary in space as time increases. 

In the simplest, most idealized case of light interacting with a $\Lambda$-configuration optical medium and in the absence of any initial medium polarization excitation, we have predicted a particularly transparent version of polarization and/or color switching~\cite{Maimistov85,byrne03}.   In particular,  if one of the two energetically-lower levels is initially more populated along the medium sample than the other,  the electric-field polarization or color of any pulse propagating through this medium switches until the pulse interacts with the medium only through the transition between the excited upper level and the less populated of the two lower levels.  In the presence of a non-vanishing initial polarization $\rho_0$, the system instead asymptotically approaches a specific combination of the two transitions, whose details depend on the value of $\rho_0$~\cite{katie07}.   Further simulations aimed at drawing analogies and comparisons between these idealized results and the dynamics of the light-metamaterial interaction investigated here will be conducted in future investigations.

\begin{acknowledgments}   A.O.K. was partially supported by University of New Mexico RAC grant \#11-19, FTP S\&SPPIR, NSh-6885.2010.2, and program ``Nonlinear Dynamics." K.E.R. was partially supported by NSF grant 
DMS-0636358. G.K. was partially supported by NSF grant
DMS-1009453. A.I.M. was partially supported by ARO-MURI award 50342-PH-MUR. I.R.G. was partially supported by NSF grant
DMS-0509589, FTP S\&SPPIR, ARO-MURI award 50342-PH-MUR and State of
Arizona (Proposition 301). 
\end{acknowledgments}

%\bibliographystyle{osajnl}
%\bibliography{mybib,optics}

\begin{thebibliography}{10}
\newcommand{\enquote}[1]{``#1''}

\bibitem{MMCE:MMCE20634}
F.~Bilotti and L.~Sevgi, \enquote{Metamaterials: Definitions, properties,
  applications, and {FDTD}-based modeling and simulation (invited paper),}
  International Journal of RF and Microwave Computer-Aided Engineering
  \textbf{22}, 422--438 (2012).

\bibitem{schurig06}
D.~Schurig, J.~J. Mock, B.~J. Justice, S.~A. Cummer, J.~B. Pendry, A.~F. Starr,
  and D.~R. Smith, \enquote{Metamaterial electromagnetic cloak at microwave
  frequencies,} Science \textbf{314}, 977--980 (2006).

\bibitem{cai06}
W.~Cai, U.~K. Chettiar, A.~V. Kildishev, and V.~M. Shalaev, \enquote{Optical
  cloaking with metamaterials,} Nature Photonics \textbf{1}, 224--227 (2007).

\bibitem{zhel12}
N.~I. Zheludev and Y.~S. Kivshar, \enquote{From metamaterials to metadevices,}
  Nature Materials \textbf{11}, 917--924 (2012).

\bibitem{smith00}
D.~R. Smith, W.~J. Padilla, D.~C. Vier, S.~C. Nemat-Nasser, and S.~Schultz,
  \enquote{Composite medium with simultaneously negative permeability and
  permittivity,} Phys. Rev. Lett. \textbf{84}, 4184--4187 (2000).

\bibitem{shelby01}
R.~A. Shelby, D.~R. Smith, and S.~Schultz, \enquote{Experimental verification
  of a negative index of refraction,} Science \textbf{292}, 77--79 (2001).

\bibitem{smith04}
D.~R. Smith, J.~B. Pendry, and M.~C.~K. Wiltshire, \enquote{Metamaterials and
  negative refractive index,} Science \textbf{305}, 788--792 (2004).

\bibitem{shalaev05}
V.~M. Shalaev, W.~Cai, U.~K. Chettiar, H.~Yuan, A.~K. Sarychev, V.~P. Drachev,
  and A.~V. Kildishev, \enquote{Negative index of refraction in optical
  metamaterials,} Opt. Lett. \textbf{30}, 3356--3358 (2005).

\bibitem{zhang05}
S.~Zhang, W.~Fan, N.~C. Panoiu, K.~J. Malloy, R.~M. Osgood, and S.~R.~J.
  Brueck, \enquote{Demonstration of near-infrared negative-index materials,}
  Phys. Rev. Lett. \textbf{95}, 137404--4 (2003).

\bibitem{shalaev07}
V.~M. Shalaev, \enquote{Optical negative-index metamaterials,} Nature Photonics
  \textbf{1}, 41--47 (2007).

\bibitem{veselago68}
V.~G. Veselago, \enquote{Electrodynamics of substances with simultaneously
  negative values of sigma and mu,} Soviet Physics Uspekhi \textbf{10},
  509--514 (1968).

\bibitem{pendry00}
J.~B. Pendry, \enquote{Negative refraction makes a perfect lens,} Phys. Rev.
  Lett \textbf{85}, 3966--3969 (2000).

\bibitem{podolskiy05}
V.~A. Podolskiy and E.~E. Narimanov, \enquote{Near-sighted superlens,} Opt.
  Lett. \textbf{30}, 75--77 (2005).

\bibitem{lu03}
J.~Lu, T.~M. Grzegorczyk, Y.~Zhang, J.~Pacheco~Jr., B.-I. Wu, J.~A. Kong, and
  M.~Chen, \enquote{\v {C}erenkov radiation in materials with negative
  permittivity and permeability,} Opt. Express \textbf{11}, 723--734 (2003).

\bibitem{agranovich04}
V.~M. Agranovich, Y.~R. Shen, R.~H. Baughman, and A.~A. Zakhidov,
  \enquote{Linear and nonlinear wave propagation in negative refraction
  metamaterials,} Phys. Rev. B \textbf{69}, 165112--7 (2004).

\bibitem{mandelshtam45}
L.~I. Mandelshtam, \enquote{Group velocity in crystalline arrays,} Zh. Eksp.
  Teor. Fiz. \textbf{15}, 475--478 (1945).

\bibitem{veselago67}
V.~G. Veselago, \enquote{Properties of materials having simultaneously negative
  values of dielectric ($\xi$) and magnetic ($\mu$) susceptibilities,} Soviet
  Physics Solid State, Ussr \textbf{8}, 2854 (1967).

\bibitem{zharov03}
A.~A. Zharov, I.~V. Shadrivov, and Y.~S. Kivshar, \enquote{Nonlinear properties
  of left-handed metamaterials,} Phys. Rev. Lett. \textbf{91}, 037401--4
  (2003).

\bibitem{scalora05}
M.~Scalora, M.~Syrchin, N.~Akozbek, E.~Poliakov, G.~D'{A}guanno, N.~Mattiucci,
  M.~Bloemer, and A.~Zheltikov, \enquote{Generalized nonlinear {S}chrodinger
  equation for dispersive susceptibility and permeability: {A}pplications to
  negative index materials,} Phys. Rev. Lett. \textbf{95}, 013902--4 (2005).

\bibitem{lazarides05}
N.~Lazarides and G.~P. Tsironis, \enquote{Coupled nonlinear {S}chrodinger field
  equations for electromagnetic wave propagation in nonlinear left-handed
  materials,} Phys. Rev. E \textbf{71}, 036614 (2005).

\bibitem{popov06}
A.~K. Popov and V.~M. Shalaev, \enquote{Negative-index metamaterials:
  Second-harmonic generation, manley-rowe relations and parametric
  amplification,} Appl. Phys. B. \textbf{84}, 131 (2006).

\bibitem{gabitov06}
I.~R. Gabitov, R.~Indik, N.~M. Litchinitser, A.~I. Maimistov, V.~M. Shalaev,
  and J.~Soneson, \enquote{Double-resonant optical materials with embedded
  metal nanostructures,} J. Opt. Soc. Am. \textbf{B 23}, 535--542 (2006).

\bibitem{litchinitser07b}
N.~M. Litchinitser, I.~R. Gabitov, A.~I. Maimistov, and V.~M. Shalaev,
  \enquote{Effect of an optical negative index thin film on optical
  bistability,} Opt. Lett. \textbf{32}, 151--153 (2007).

\bibitem{litchinitser07c}
N.~M. Litchinitser, I.~R. Gabitov, and Maimistov, \enquote{Optical bistability
  in a nonlinear optical coupler with a negative index channel,} Physical
  Review Letters \textbf{99}, 113902 (2007).

\bibitem{maimistov07a}
A.~I. Maimistov and I.~R. Gabitov, \enquote{Nonlinear optical effects in
  artificial materials,} European Phys. J. -- Special Topics \textbf{147},
  265--286 (2007).

\bibitem{shadrivov04}
I.~V. Shadrivov, A.~A. Sukhorukov, Y.~S. Kivshar, A.~A. Zharov, A.~D. Boardman,
  and P.~Egan, \enquote{Nonlinear surface waves in left-handed materials,}
  Phys. Rev. E \textbf{69}, 016617--9 (2004).

\bibitem{gabitov07b}
I.~R. Gabitov, A.~I. Maimistov, A.~Korotkevich, and J.~B. McMahon,
  \enquote{Solitary waves in plasmonic bragg gratings,} Applied Physics A
  \textbf{89}, 277--281 (2007).

\bibitem{konopnicki81}
M.~J. Konopnicki and J.~H. Eberly, \enquote{Simultaneous propagation of short
  different-wavelength optical pulses,} Phys. Rev. A \textbf{24}, 2567--2583
  (1981).

\bibitem{Pendry19112004}
J.~B. Pendry, \enquote{A chiral route to negative refraction,} Science
  \textbf{306}, 1353--1355 (2004).

\bibitem{PhysRevE.69.026602}
T.~G. Mackay and A.~Lakhtakia, \enquote{Plane waves with negative phase
  velocity in {F}araday chiral mediums,} Phys. Rev. E \textbf{69}, 026602
  (2004).

\bibitem{Maimistov07}
A.~I. Maimistov, I.~R. Gabitov, and E.~V. Kazantseva, \enquote{Quadratic
  solitons in negative refractive index medium,} Optics and Spectroscopy
  \textbf{102}, 90--97 (2007).

\bibitem{MaimistovGabitovOS08}
A.~Maimistov, I.~Gabitov, and N.~Litchinitser, \enquote{Solitary waves in
  nonlinear oppositely directed coupler,} Optics and Spectroscopy \textbf{104},
  253--257 (2008).

\bibitem{Podolskiy05b}
V.~A. Podolskiy and E.~E. Narimanov, \enquote{Strongly anisotropic waveguide as
  a nonmagnetic left-handed system,} Phys. Rev. B \textbf{71}, 201101(R)
  (2005).

\bibitem{Naik05062012}
G.~V. Naik, J.~Liu, A.~V. Kildishev, V.~M. Shalaev, and A.~Boltasseva,
  \enquote{Demonstration of {Al:ZnO} as a plasmonic component for near-infrared
  metamaterials,} Proceedings of the National Academy of Sciences \textbf{109},
  8834--8838 (2012).

\bibitem{xiao10}
S.~Xiao, V.~P. Drachev, A.~V. Kildishev, X.~Ni, U.~K. Chettiar, H.-K. Yuan, and
  V.~M. Shalaev, \enquote{Loss-free and active optical negative-index
  metamaterials,} Nature \textbf{466}, 735--738 (2010).

\bibitem{allen87}
L.~Allen and J.~H. Eberly, \emph{Optical Resonance and Two-Level Atoms} (Dover,
  New York, 1987).

\bibitem{ziolkowski}
R.~W. Ziolkowski and E.~Heyman, \enquote{Wave propagation in media haveing
  negative permittivity and permeability,} Phy. Rev. E \textbf{056625} (2001).

\bibitem{gabkenmaim10}
I.~R. Gabitov, B.~Kennedy, and A.~I. Maimistov, \enquote{Coherent amplification
  of optical pulses in metamaterials,} IEEE J. of Selected Topics in Quantum
  Electronics \textbf{16}, 401--409 (2010).

\bibitem{panina02}
L.~Panina, A.~N. Grigorenko, and D.~P. Makhnovskiy, \enquote{Optomagnetic
  composite medium with conducting nanoelements,} Phys. Rev. B \textbf{66},
  155411 (2002).

\bibitem{smith06}
D.~R. Smith and J.~B. Pendry, \enquote{Homogenization of metamaterials by field
  averaging,} J. Opt. Soc. Am. \textbf{23}, 391--403 (2006).

\bibitem{klar06}
T.~A. Klar, A.~V. Kildishev, V.~P. Drachev, and V.~M. Shalaev,
  \enquote{Negative-index metamaterials: Going optical,} IEEE J. of Selected
  Topics in Quantum Electronics \textbf{12}, 1106--1115 (2006).

\bibitem{Pendry99}
J.~B. Pendry, A.~J. Holden, D.~J. Robbins, and W.~J. Stewart,
  \enquote{Magnetism from conductors and enhanced nonlinear phenomena,} IEEE
  Trans. Microwave Theory Tech. \textbf{47}, 2075--2084 (1999).

\bibitem{PhysRevE.65.036622}
P.~Marko\ifmmode~\check{s}\else \v{s}\fi{} and C.~M. Soukoulis,
  \enquote{Numerical studies of left-handed materials and arrays of split ring
  resonators,} Phys. Rev. E \textbf{65}, 036622 (2002).

\bibitem{cohen}
C.~Cohen-Tannoudji, B.~Diu, and F.~Laloe, \emph{Quantum Mechanics} (Wiley, New
  York, 1977).

\bibitem{Maimistov85}
A.~I. Maimistov and Y.~M. Sklyarov, \enquote{Coherent interaction of light
  pulses with a three-level medium,} Optics and Spectroscopy \textbf{59},
  459--461 (1985).

\bibitem{byrne03}
J.~A. Byrne, I.~R. Gabitov, and G.~Kova\v{c}i\v{c}, \enquote{Polarization
  switching of light interacting with a degenerate two-level optical medium,}
  Physica D \textbf{186}, 69--92 (2003).

\bibitem{katie07}
K.~A. Newhall, \enquote{Three level atom interaction with external electric
  field,}  (2007). Notes.

\end{thebibliography}

\end{document}